\documentclass[prb,onecolumn,superscriptaddress,amsmath,amsfont,amssymbs,noshowpacs]{revtex4}
\usepackage[version=3]{mhchem}
\usepackage{graphicx}
\usepackage{multirow}
\usepackage{textcomp}
\usepackage{hyperref}
\usepackage[textsize=tiny]{todonotes}
\setcitestyle{super,open={},close={}}

\usepackage{color}
\usepackage{soul}

\begin{document}

\title{Metal-free perovskites for non-linear optical materials}

\author{Thomas W.\ Kasel}
\affiliation{Department of Chemistry, University of Oregon Eugene, Oregon, 97403}
\affiliation{\textcolor{black}{contributed equally}}

\author{Zeyu Deng}
\affiliation{Department of Materials Science and Engineering, National University of Singapore, Singapore 117575, Singapore}
\affiliation{\textcolor{black}{contributed equally}}

\author{Austin M.\ Mroz}
\affiliation{Department of Chemistry, University of Oregon Eugene, Oregon, 97403}
\affiliation{\textcolor{black}{contributed equally}}

\author{Christopher H. Hendon}
\email{chendon@uoregon.edu}
\affiliation{Department of Chemistry, University of Oregon Eugene, Oregon, 97403}

\author{Keith T. Butler}
\email{keith.butler@stfc.ac.uk}
\affiliation{ISIS Neutron and Muon Source, Rutherford Appleton Laboratory, Harwell Campus, Harwell, United Kingdom}

\author{Pieremanuele Canepa}
\email{pcanepa@nus.edu.sg}
\affiliation{Department of Materials Science and Engineering, National University of Singapore, Singapore 117575, Singapore}



\begin{abstract}
We identify the existence of nonlinear optical (NLO) activity in a number of novel $ABX_3$-type metal-free perovskites, where $A$ is a highly tuneable organic cation, $B$ is a NH$_4$ cation and $X$ a halide anion. Through systematic first-principles calculations, we  identify  important  trends to chart the second-harmonic generation of this class of materials. We study three perovskites MDABCO-NH$_4$I$_3$, CNDABCO-NH$_4$I$_3$ and ODABCO-NH$_4$I$_3$ for use as deep-UV second-harmonic generation materials. We identify the role of the dipole moment imparted by the organic group on the $A$ cation as an important parameter to tune the NLO properties of these materials. We apply this knowledge functionalising  the organic group DABCO with the highly polar cyanide CN$^-$ group, and we demonstrate a significant improvement of the NLO response in this family of materials.   These  findings can accelerate the application of metal free perovskites as inexpensive, non-toxic, earth-abundant materials for the next generation of optical communication applications.
\end{abstract}
\maketitle


\section{Introduction}


Light has been utilised as a communication device for many centuries. Recently, non-linear optics (NLO) and second harmonic generation (SHG) have been at the heart of several technological revolutions. With the advent of  the internet, conveying information and data by means of fibre-optics and lasers have transformed telecommunications. The development of fiberoptic devices with increased performance has fuelled a surge of interest in the development of materials with ever-increasing data-transfer capabilities.\cite{Franken1961,Stucky1989,Eaton1991,Bredas1994,Halasyamani1998,Schneider2004,Wu2013,Rondinelli2015,Tran2016,Halasyamani2018}   Materials, such as LiNbO$_3$ and LiTaO$_3$ with  NLO and SHG  are at the core of high-speed electro-optic modulator devices, significantly boosting the transmission capacities of the telecommunication infrastructure ($\sim$~10 Gbit~s$^{-1}$).\cite{Franken1961,Cyranoski2009} In parallel NLO materials for deep-UV lasers are being used in semiconductor manufacturing, photolithography, laser systems, and advanced instrument development.

The demand for apparatus with increased performance requires the development of novel inexpensive NLO materials and every year the electronic and telecommunication industries demands the production of $\sim$40,000 tons of LiNbO$_3$.  The soaring costs of lithium and niobium\cite{metalary,Olivetti2017} and thus LiNbO$_3$, requires the development of the new generation of NLO materials relying on more earth-abundant elements.\cite{Olson2000,Chung2013,Goud2018} This is further aggravated by the fact that LiNbO$_3$ melts incongruently and the manufacturing of congruently lithium niobate single crystals requires cooling of Li-poor nonstoichiometric melts of LiNbO$_3$,\cite{Wilkinson1993} adding additional production costs.  

Meanwhile, for deep-UV applications Tran \emph{et al.}\cite{Tran2016} have shown that  only a handful of materials, i.e., KBe$_2$BO$_3$F$_2$, RbBe$_2$BO$_3$F$_2$ and  CsBe$_2$BO$_3$F$_2$,  sharing similar structural features  currently fulfil the desired requirements. However, these  materials contain toxic Be, whose usage is prohibited in many countries. Therefore, materials with better NLO characteristics are required to supplant deep-UV lasers.\cite{Halasyamani2018}  As identified by Halasyamani and Rondinelli,\cite{Halasyamani2018} NLO materials for application in deep-UV lasers must fulfil a minimal number of requirements: \emph{i}) the atoms in the material should present a noncentrosymmetric arrangement, \emph{ii}) the material absorbs light in the deep-UV spectrum (i.e., absorption wavelengths $\geq$~175~nm), and \emph{iii})  the material should respect the phase-matching criteria discussed in detail within the manuscript.

One promising route towards sustainable NLO materials lies in  organic materials, which have shown promising  SHG properties.{\cite{Zyss1981,Eaton1991,Prasad1991,Kaino1993,Pan1996,Albert1998,Jiang1999,Kuo2001,Bosshard2002,Evans2002,Marder2006,Wang2011}} However, to date organic NLO suffer from thermal instability and difficult fabrication. In contrast to typical organic-based NLO, metal-free perovskites have been shown to be structurally stable beyond  200 $^\circ$C.{\cite{Ye2018}}

In this study we present an in-depth analysis of the NLO properties of a new class of materials with ferroelectric response  termed metal-free perovskites developed by Ye and co-workers.\cite{Ye2018} We use first-principles calculations, based on density functional theory (DFT) to chart the optical and NLO properties of a these novel metal-free perovskites. In contrast to typical inorganic perovskite $A^{2+}B^{4+}O^{2-}_3$ where $A$ and $B$ are metal cations (e.g., CaTiO$_3$), in metal-free perovskites the $A$ and $B$ cations are replaced by organic units.  The rich choice of  organic units introduces the possibility of tailoring the functional properties of such metal-free perovskites, while being easy-to-synthesise, affordable and non-toxic. Note that NLO materials with  inorganic perovskite structures exist (e.g., K$_3$B$_6$O$_{10}$Cl) which are more closely related to the metal-free perovskites than organic-based NLO materials.\cite{Wu2011} 

We investigate such metal-free perovskites as NLO materials for SHG applications, and verify whether these perovskites are suitable as deep-UV NLO materials. Our  findings reveal that metal-free perovskites, based on the organic moieties N-methyl-N-diazabicyclo[2.2.2]octonium (MDABCO$^{2+}$) and N-hydroxy-N-diazabicyclo[2.2.2]octonium (ODABCO$^{2+}$), posses SHG response, with magnitudes similar to  some inorganic contenders, such as KBe$_2$BO$_3$F$_2$, RbBe$_2$BO$_3$F$_2$ and  CsBe$_2$BO$_3$F$_2$.  We demonstrate that organic groups presenting intrinsic dipole moments can contribute positively to the ferroelectric response of the material, and provide an increased NLO response. 

On the basis of our predictions, we provide guidelines to improve the SHG response of the metal-free perovskites, by tailoring the structural features of the organic cations. Following this principle, we extend the computational search to new structures. For example, we show the case where the hydroxy-group in ODABCO$^{2+}$ ($A$-cation) is effectively replaced by a cyanide CN$^-$ group with increased polarity, which provides remarkable NLO response. While preliminary observations\cite{Ye2018} showed experimentally  SHG response in MDABCO-NH$_4$I$_3$, we also identify ODABCO-NH$_4$I$_3$ and CNDABCO-NH$_4$I$_3$ as superior SHG materials, and their properties should be carefully verified experimentally.  


\section{Results}


\subsection{Structure of ferroelectric metal-free perovskites}

\begin{figure}[h!]
\begin{center}
{
\includegraphics[width=\columnwidth]{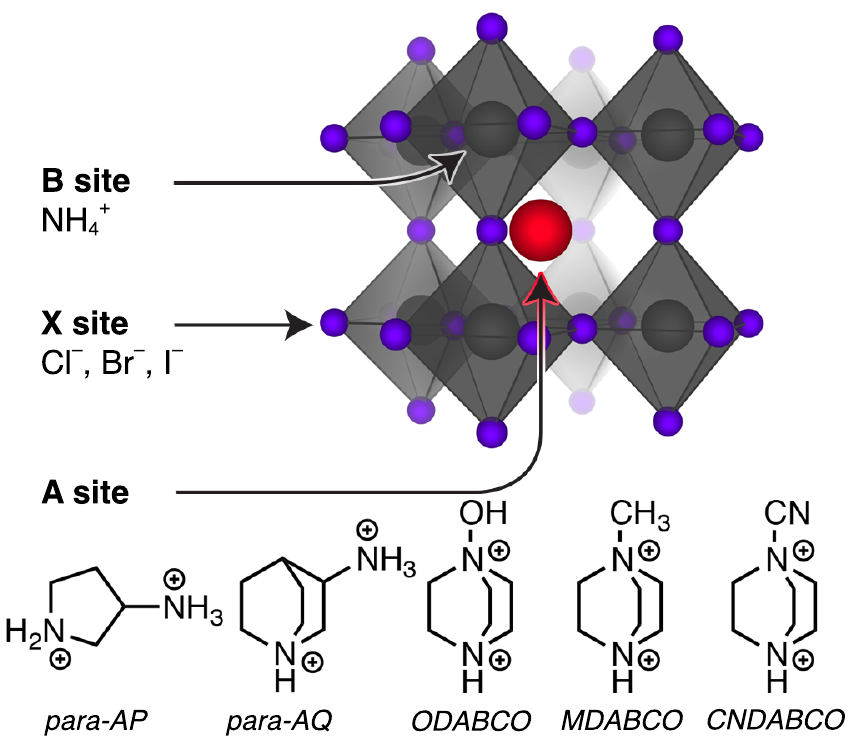}}
\caption{\label{fig:structures} Polyhedra representation of a metal-free $ABX_3$-type perovskite, where $A$  can be one of the four organic cations (also shown by the red ball), $B$ is NH$_4^+$ (grey ball) and $X$ the halide ion, i.e. Cl$^-$, Br$^-$ and I$^-$ (purple balls).  The nature of the $A$ cation is also shown. 
}
\end{center}
\end{figure}

The first point when considering the suitability of a material for NLO applications is the crystal structure. All materials reported 
herein and shown in Figure~\ref{fig:structures} feature the typical perovskite $ABX_3$ structure (where $A$ and $B$ are cations and $X$ are anions), with $BX_3$ corner-sharing octahedra and charge-balancing $A$ cations in the cavities of the framework. The structures investigated have NH$_4^+$ as $B$ sites and halide X$^-$ sites, as seen in Figure~\ref{fig:structures}. A series of different $A$ site divalent molecular cations are considered (i)  N-methyl-N-diazabicyclo$[$2.2.2$]$octonium (MDABCO$^{2+}$), (ii) M-hydroxy-N-diazabicyclo$[$2.2.2$]$octonium (ODABCO$^{2+}$), (iii) R-3-ammonioquinuclidinium (R-3AQ$^{2+}$), and (iv) S-3-ammoniopyrrolidinium (S-3AP$^{2+}$). Starting from the ODABCO-NH$_4$X structures, we also replaced the hydroxy group with a more polar cyanide CN$^-$ group forming a new metal-free perovskite termed  CNDABCO-NH$_4$X$_3$ .
In studying the halide chemical space of these metal-free perovskites, we proposed three new structures, including MDABCO-NH$_4$Cl$_3$, ODABCO-NH$_4$I$_3$ and S-3AP-NH$_4$I$_3$ that were not identified by Ye \emph{et al.}\cite{Ye2018}  In the case of ODABCO-NH$_4$I$_3$ and CNDABCO-NH$_4$X$_3$ we assumed a $R3$ space group. However, we also computed the Iodine-based polymorph with $Pca2_1$ space group (as in ODABCO-NH$_4$Cl$_3$), which  converted to the $R3$ polymorph. 

To be NLO active a material must be non-centrosymmetric. As shown in Table~S1 and Table~S2 of the Supporting Information (SI), the materials considered here fall into three 
spacegroups, $R3$, $Pca2_1$ and $P2_1$, all of which are non-centrosymmetric. 

In addition,  metal-free perovskites,  and inorganic materials, such as LiNbO$_3$  posses an intrinsic polarization, which is typical of  ferroelectric materials. LiNbO$_3$ charts among the most popular of NLO materials with a  polarization of $\sim$~70-75 $\mu$C/cm$^2$.\cite{Veithen2002} The presence of dipolar molecules and the off-centring of the $B$-site cation in the metal-free perovskites indicates the existence of an intrinsic polarization, which for MDABCO-NH$_4$I$_3$ has been measured to be $\sim$~19~$\mu$C/cm$^2$.\cite{Ye2018} 

\subsection{Birefringence activity of Metal-free Perovskites}
\noindent Having established the crystallographic criteria for NLO materials, we now consider the optical properties that must be satisfied for SHG. The birefringence is the maximum difference  refractive indices  ($n$) which depends on the propagation directions  of light in the crystal at a fixed frequency ($\omega$)  (as in Eq.~\ref{eq:spreadn}).
\begin{equation}
\label{eq:spreadn}
\Delta n = n_{\rm max}({\rm \omega}) - n_{\rm min}({\rm \omega})
\end{equation}
SHG occurs  efficiently whenever the refractive index $n$ of the 2$^{nd}$ harmonic is equal (or close) to that of the generating wave at half-frequency, i.e., $n(2{\rm \omega}) =n({\rm \omega})$  ---this condition is termed phase-matching. Phase-matching requires a suitable 
range of frequencies with refractive indexes, $n_{\rm max}({\rm \omega}) - n_{\rm min}(2{\rm \omega}) > 0$, with $n_{\rm min}$ and 
$n_{\rm max}$ the lowest and the largest refractive index in the spread.\cite{Zhang2017}  

If $\Delta n$ of a material is too small the  phase-matching condition for SHG will not occur. If $\Delta n$ is too large 
the material will exhibit spatial beam walk-off, where the intensity distribution of the wave drifts away from the direction of 
propagation resulting in reduced SHG intensity. In general a moderate birefringence ($\Delta n \approx$ 0.07) is desired. 

In uniaxial systems, such as LiNbO$_3$ and the MDABCO-NH$_4$X$_3$ systems the direction of ordinary and extraordinary rays occurs along the optical axis, which lays along the highest symmetry axis. 

\begin{figure}[h]
\begin{center}
{
\includegraphics[width=\columnwidth]{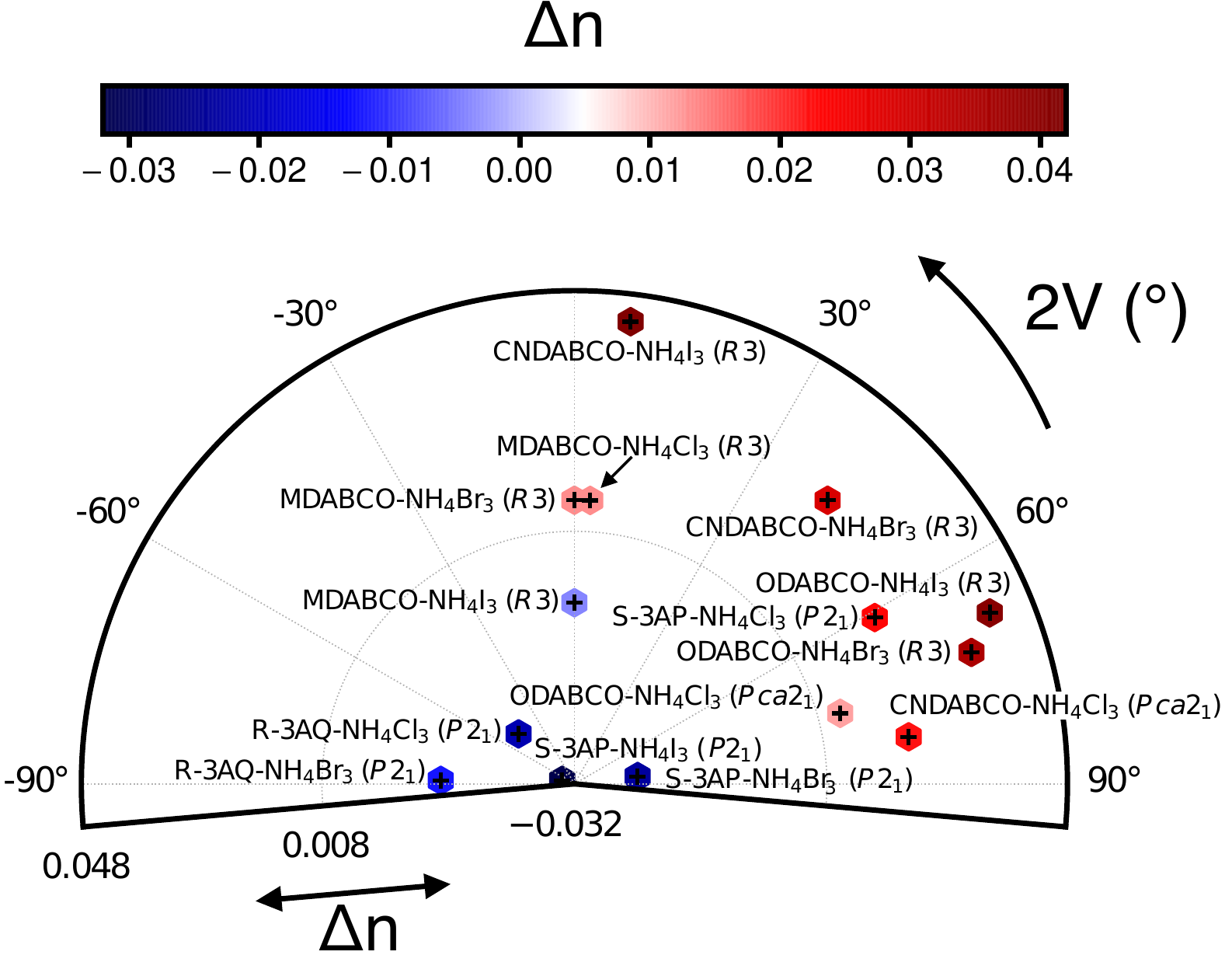}}
\caption{\label{fig:nindexes} Computed birefringence $\Delta n$ and 2V angle of metal-free perovskites at their experimental volume and lattice constants. $\Delta n$ is plotted on the radial axis and using the colour-bar, with red indicating positive $\Delta n$. Values of $n$ are given in Table S3 of the SI. The 2V angles provide information about the nature of the birefringence in biaxial crystals. 
}
\end{center}
\end{figure}

In contrast, biaxial systems have multiple optic axes, whose directions of propagation depend on the diffraction index and measured by the 2V acute angle (see below).\cite{Zhang2017} Therefore, the 2V angle provides information about the nature of the birefringence in biaxial crystals.   In general, orthorhombic,  monoclinic and triclinic systems are biaxial crystals. 

Figure~\ref{fig:nindexes}  plots $\Delta n$ (obtained from the dielectric constants in the static regime) and the 2V acute angle.  
From Figure~\ref{fig:nindexes}, we observe that all the systems considered here have $\Delta n$  ranging between values of $-$0.03 and +0.05, with CNDABCO, ODABCO, S-3AP and R-3AQ-based perovskites charting among the highest in magnitude. The $R3$ materials, which are uniaxial NLO systems, all exhibit positive birefringence.  In contrast, the $Pca2_1$ and $P2_1$ systems, which are biaxial NLO materials, display both positive and negative birefringence. A closer look at Figure~\ref{fig:nindexes} shows that with the exception of  MDABCO-based perovskites and S-3AP-NH$_4$I$_3$ (whose 2V~$\sim$~0), the remaining materials all show a rather complex biaxial response.

\subsection{NLO response in Metal-free Perovskites}

We now  discuss the NLO response of the metal-free perovskites. Among the prerequisites for SHG the materials under investigation should  display a non-negligible value of the second order dielectric tensor,  $\chi ^{(2)}$.

$\chi^{(2)} _{ijk} = 2d_{ijk}$  is a third-rank tensor as defined in Eq.~{\ref{eq:beta}}, and therefore can be difficult to analyse intuitively. The $i$, $j$ and $k$ components of the $d$ tensor identify the directions of the applied electric fields (i.e., $j$ and $k$)  of the incident radiation and the polarization of the generated second harmonic  (i.e., $i$), respectively, (see Eq.~{\ref{eq:beta}}). In SHG experiments, the directions of the electric fields of incident radiations have frequencies $\omega_1$  and $\omega_2$ (with $\omega_1 = \omega_2$ in SHG) and the second harmonic wave with frequency $\omega_3 = \omega_1 +\omega_2$.  A complete discussion of the  $\chi^{(2)}$ and $d$ tensors and their dependence can be found in Ref.~\citenum{Boyd2008}.   

A useful proxy for presenting the $ijk$ component of the static $\chi^{(2)}$ tensor is the norm of the $d_{ijk}$ components of Table~{\ref{tb:chi}} and  Eq.~\ref{eq:dnorm}: 
\begin{equation}
\label{eq:dnorm}
\left||d_\mathrm{norm} \right||= \sqrt{ \sum _{i j k} d_{ijk}^2} 
\end{equation}
$d_\mathrm{norm}$ has been utilised previously{\cite{Zyss1994,Goud2018}} and is more effective of a simple geometric mean of the $d_{ijk}$ components, which can assume values approaching numerical zeros given the oscillations in sign of the $d_{ijk}$ values (see Table~{\ref{tb:chi}}). We note that the values $d_\mathrm{norm}$ are by definition larger in magnitude of any component $d_{ijk}$, but here $d_\mathrm{norm}$ serves us simply to compare each materials investigated on the same footing. 
\begin{table*}[h]
\caption{\label{tb:chi} Computed static d$_{ijk}$ components of the $ \chi^{(2)}$ tensor and their norm d$_{\rm norm}$ (in pm/V)  of the metal-free perovskites at the experimental volumes. Spg. is the space group.  }
\small
\begin{center}
\begin{tabular}{lcccccccccccc}
\hline \hline \\[-2.4mm]
X$^-$    & Spg.  &  $d_\mathrm{xxx}$       & $d_\mathrm{xxy}$       & $d_\mathrm{xxz}$  & $d_\mathrm{xyy}$       & $d_\mathrm{xyz}$       & $d_\mathrm{xzz}$  & $d_\mathrm{yyy}$       & $d_\mathrm{yyz}$  & $d_\mathrm{yzz}$       & $d_\mathrm{zzz}$       & $\left|| d_\mathrm{norm} \right||$   \\[0.5mm]
\hline\\[-2.4mm]
\multicolumn{13}{c}{MDABCO$-$NH$_4$X$_3$} \\
\hline
Cl   & $R3$  & 0.140& $-$0.410& 0.048 & $-$0.140& 0.000  & 0.000& 0.410& 0.048 & 0.000& $-$0.490    & 0.79  \\
Br   & $R3$  & 0.230 & $-$0.620& $-$0.058& $-$0.230& 0.000 & 0.000& 0.620& $-$0.058 & 0.000  & $-$0.580& 1.10  \\
I    & $R3$  & $-$0.170& 1.200& 0.430& 0.170& 0.000& 0.000& $-$1.200 & 0.430& 0.000& 0.021 & 1.80 \\
\hline \\[-2.4mm] 
\multicolumn{13}{c}{ODABCO$-$NH$_4$X$_3$} \\
\hline
Cl & $Pca2_1$ & 0.000 & 0.000 & 0.230& 0.000& 0.000& 0.000& 0.000& $-$0.076& 0.000& $-$0.076 & 0.25 \\
Br & $R3$     & 0.019 & 0.510& $-$0.320& 0.150& $-$0.290& $-$0.069& 0.099 & $-$0.100& $-$0.280& $-$0.230 & 0.79\\
I   & $R3$    & 0.270& 0.620& $-$0.670& $-$0.190& $-$0.400& 0.100& $-$0.840 & $-$0.490 & $-$0.280 & 0.007   & 1.50  \\
\hline \\[-2.4mm] 
\multicolumn{13}{c}{CNDABCO$-$NH$_4$X$_3$} \\
\hline
Cl & $Pca2_1$ & 0.000& 0.000& 0.650 & 0.000 & 0.000 & 0.000 & 0.000 & $-$0.160 & 0.000 & 0.075 & 0.68 \\
Br  & $R3$    & $-$0.450& 0.850 & 0.260 & 0.230&  0.120 & 0.092 & $-$0.880 & 0.350 & 0.071 & $-$0.590& 1.50\\
I  & $R3$     & $-$0.450 & 1.310 & 0.760 & 0.450 & $-$0.001& 0.0006  & $-$1.310 & 0.760 & 0.004  & $-$0.014 & 2.20  \\
\hline \\[-1.5mm] 
\multicolumn{13}{c}{R$-$3AQ$-$NH$_4$X$_3$} \\
\hline
Cl   & $P2_1$       & 0.000 & 0.091 & 0.000& 0.000  & 0.068 & 0.000& 0.012 1 & 0.000& $-$0.160& 0.000& 0.19\\
Br  & $P2_1$     & 0.000 & $-$0.075 & 0.000& 0.000 & 0.004   & 0.000 & $-$0.320& 0.000  & 0.400 & 0.000& 0.52\\
\hline \\ [-2.4mm]
\multicolumn{13}{c}{S$-$3AP$-$NH$_4$X$_3$} \\
\hline
Cl  & $P2_1$       & 0.000  & $-$0.025 & 0.000 & 0.000 & $-$0.041 & 0.000 & $-$0.160 & 0.000 & $-$0.150& 0.000& 0.22 \\
Br & $P2_1$       & 0.000& $-$0.029& 0.000& 0.000 & $-$0.055 & 0.000& $-$0.120& 0.000 & $-$0.048 & 0.000& 0.14 \\
I & $P2_1$       & 0.000& $-$0.210& 0.000& 0.000 & $-$0.075 & 0.000& $-$0.140 & 0.000 & 0.029 & 0.000& 0.26  \\ 
\hline \\[-2.4mm] 
\multicolumn{13}{c}{LiNbO$_3$} \\
\hline
   ---          & $R3c$       & 0.170&  0.000  & 5.000& 0.170 & 0.000& 0.000& 0.000 & 5.000& 0.000& 16.450  & 17.91 \\
\hline \hline
\end{tabular}
\end{center}
\end{table*}

\begin{figure}[h!]
\begin{center}
{
\includegraphics[width=\columnwidth]{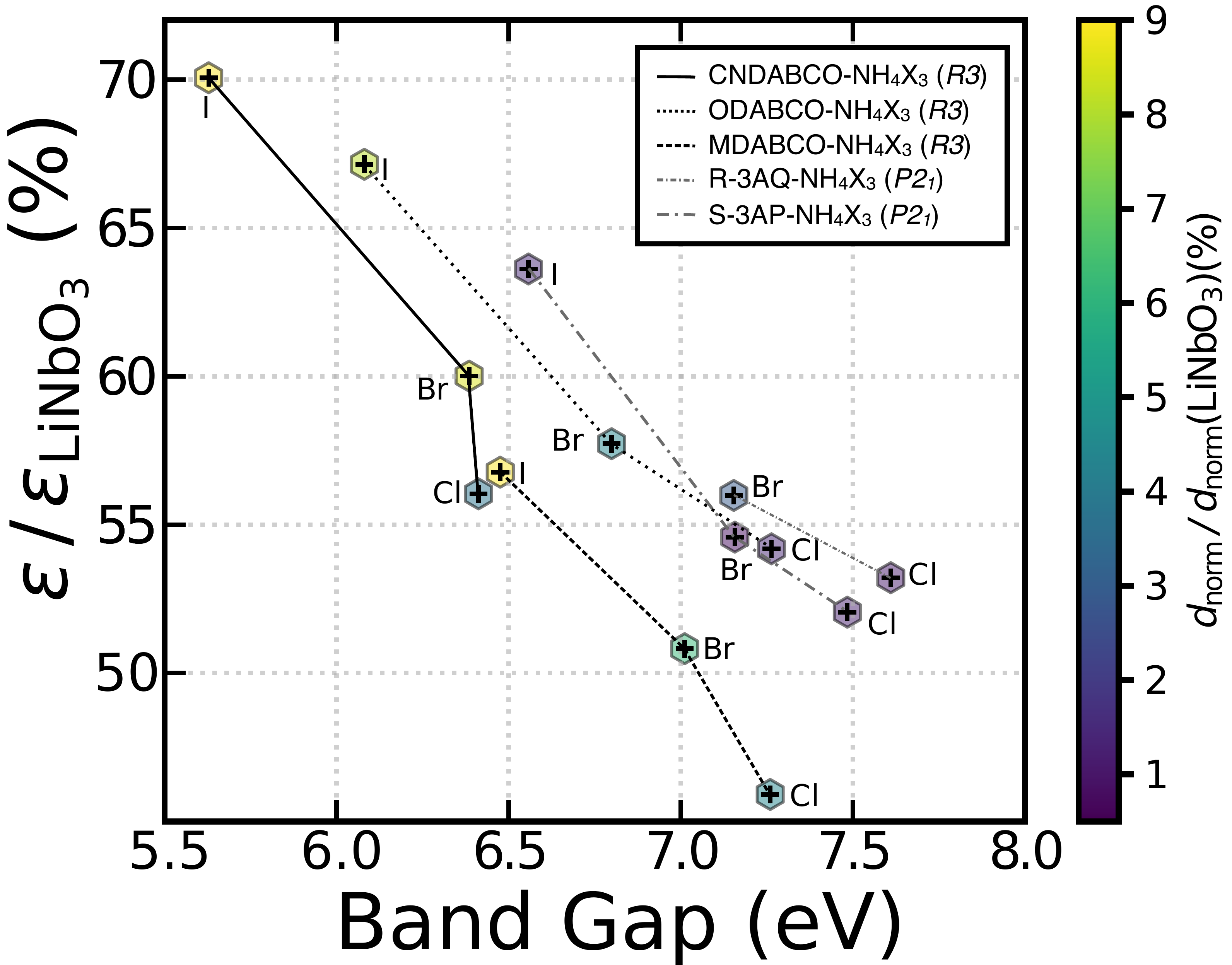}}
\caption{\label{fig:NLO}  Computed static $d_{\rm norm}$ (colour-bar) as function of the band gap (x-axis, eV) and average dielectric constant (y-axis) of the metal-free perovskites at the experimental volume. The $\epsilon$ values (in the static limit) computed for the metal-free perovskites are normalised against the LiNbO$_3$ ($\sim$ 4.39).  Similarly, the  values of $d_{\rm norm}$ are rescaled to that of LiNbO$_3$ ($\sim$~17.91~pm/V).  The space group of each structure is also reported. 
}
\end{center}
\end{figure}

Figure~\ref{fig:NLO} presents the static dielectric constant $\varepsilon$ (y-axis) as a function of the band gap (x-axis) for the materials studied, the data points are coloured according to the magnitude of $d_{\rm norm}$, which are obtained at the experimental volumes at room temperature. The values of band gap, $\varepsilon$,  $\chi^{(1)}$, $\chi^{(2)}_{ijk}$ and d$_{\rm norm}$ of these materials are found in Table S4 and Table S5. 

In general, volumes obtained after relaxation in hybrid systems containing a number of Van der Waals interactions, such as these metal-free perovskites, tend to be  highly distorted. Therefore, the main text concentrates on results obtained on structures where all the atomic coordinates are relaxed at the experimental volumes and lattice constants, whereas results on fully relaxed structures are available in the SI. For clarity and ease of comparison, the dielectric constant and the $d_{\rm norm}$ are scaled by that of LiNbO$_3$, calculated at the same level of theory (Table S4 and Table S5). 

The y-axis of Figure~\ref{fig:NLO} presents the average  dielectric  constant (in the static limit) of the metal-free perovskites rescaled to that of  LiNbO$_3$ $\sim$~4.39, which compare well with experimental data ($\sim$~4.87).\cite{Veithen2002}  The computed $d_{zzz}$ values of 16.45~pm/V (at the experimental volume of LiNbO$_3$) slightly underestimated the experimental value of ($\sim$22.0 pm/V at $\lambda \sim$ 407.2~nm) by Magel \emph{et al.}\cite{Magel1990}  DFT has been  known to systematically underestimate values of $d$ in NLO materials.\cite{Lin2014} The complete dielectric and $d$ tensors of each metal-free perovskites are reported in the SI. 

The computed band gaps of Figure~\ref{fig:NLO} are characteristic of high-gap insulators ranging between $\sim$~5.5 and 8~eV. The metal-free perovskites absorb light between 155 and 220~nm, i.e.\ the deep-UV portion of the electromagnetic spectrum, meaning that these materials fulfil another of the criteria for deep-UV SHG applications. All computed $d_{\rm norm}$ values in metal-free perovskites with the same organic A-site cation follow the order I  $>$ Br  $>$ Cl. The order of increasing $d_{\rm norm}$ moving down the halide group follows the trend in band gaps. We find that the computed $d_{\rm norm}$ of all the metal-free perovskites is a fraction (ranging between 1\% and 10\%) of the the $d_{\rm norm}$ of LiNbO$_3$ (Figure~\ref{fig:NLO}), which is used as a reference. In MDABCO-NH$_4$I$_3$, $d_{2,1}$ and $d_{2,2}$ have the largest values ($\sim$1.196 pm/V) corresponding to a $\chi^{(2)}$ of $\sim$1.23~a.u., which is in agreement with the SHG measurements of Ye~\emph{et al.},{\cite{Ye2018}} reporting a value of $\sim$1.35~a.u. at 300 K.

From these data, we identify two iodine-based structures, MDABCO-NH$_4$I$_3$ and ODABCO-NH$_4$I$_3$  as the best NLO materials across the metal-free perovskites investigated. When the OH$^-$ group in ODABCO-NH$_4$X$_3$ is replaced by a polar CN$^-$ group (see Figure~{\ref{fig:NLO}}), forming  CNDABCO-NH$_4$X$_3$, we find improved NLO properties. In CNDABCO-NH$_4$X$_3$, the computed $d_{\rm norm}$ of $\sim$2.23~pm/V outperforms all other metal-free perovskites, suggesting that CN-substituted   $A$ cations offer a promising strategy to improve the optoelectronic properties of these materials. 
 
Figure~\ref{fig:NLO} shows that both R-3AQ-NH$_4$X$_3$ and S-3AP-NH$_4$X$_3$ posses small values of $d_{\rm norm}$, covering a narrow range between $\sim$0.14 and $\sim$0.52 pm/V, with the highest value set by R-3AQ-NH$_4$Br$_3$ and the lowest by S-3AP-NH$_4$Br$_3$. Hence, both R-3AQ-NH$_4$X$_3$ and S-3AP-NH$_4$X$_3$ are expected to show low SHG behaviour and therefore, will not be considered further in this analysis.


\section{Discussion}


\begin{figure*}[!h]
\begin{center}
{
\includegraphics[width=1.0\columnwidth]{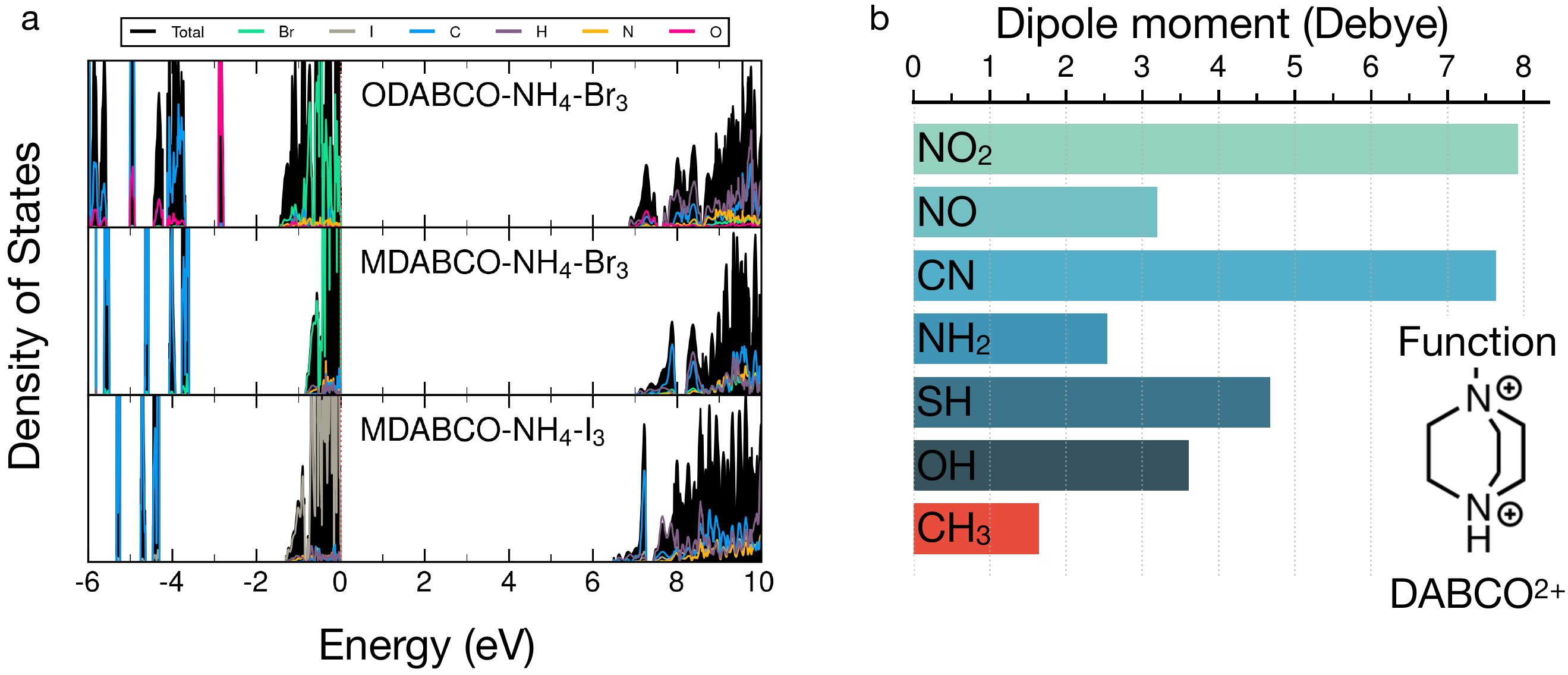}}
\caption{\label{fig:DOS}  {\bf a} Total and projected density of states of ODABCO-NH$_4$Br$_3$, MDABCO-NH$_4$Br$_3$ and MDABCO-NH$_4$I$_3$ obtained with PBESol0-D3. The red line (dash) is the Fermi energy, which is set at the top of the valence band of each material. {\bf b} Variance of  dipole moment for DABCO cations substituted with a number of organic groups (functions).
}
\end{center}
\end{figure*}

\noindent We have identified that MDABCO-NH$_4$I$_3$, ODABCO-NH$_4$I$_3$ and CNDABCO-NH$_4$I$_3$ show the largest NLO response. Our results show the largest components of the $d$ tensor in MDABCO-NH$_4$I$_3$ are $d_{xxz}=d_{yyz}\sim$0.43 pm/V and $d_{xxy}\sim$1.20~pm/V, respectively,  $d_{yyy}\sim$0.84~pm/V in ODABCO-NH$_4$I$_3$ and $d_{yyy}\sim$1.31~pm/V  in CNDABCO-NH$_4$I$_3$, which represent three promising NLO materials with good SHG. Therefore, the discussion will only focus on the subset of metal-free perovskites showing significant SHG activity, including CNDABCO-NH$_4$I$_3$, ODABCO-NH$_4$X$_3$ and  MDABCO-NH$_4$X$_3$.  

The values for SHG are similar and in most cases even larger than  the  prescribed  minimal conditions for deep-UV applications, i.e.\ $d_{xyz}\sim$0.39~pm/V (with $\lambda \sim$1.064~$\mu$m) KH$_2$PO$_4$.\cite{Nikogosyan1997,Tran2016} In combination with the calculated band gaps, our study shows that these materials are very promising for application as deep-UV NLO materials.  The composition-structure-property relationships driving NLO response in these test systems can provide rules for designing even more SHG active materials.

As seen in Figure~\ref{fig:NLO}, the primary factor influencing the size of $d_{\rm norm}$ is the magnitude of the band gap. This is a well established trend in semiconductor materials in general and is related to lower carrier concentrations in wider gap materials at finite temperatures. A correlation between band-gap and the halide species exists, we find a similar correlation in the values of the static dielectric constants vs.\ the anion species, following the trend $\varepsilon ($I$^-)>\varepsilon ($Br$^-)>\varepsilon ($Cl$^-$). This observation is consistent with the density of states of Figure~\ref{fig:DOS} of ODABCO-NH$_4$Br$_3$ (top), MDABCO-NH$_4$Br$_3$ (middle) and MDABCO-NH$_4$I$_3$ (bottom), respectively, which shows the valence band dominated by halide species and the conduction band populated by the $A$ organic cation, i.e., ODABCO$^{2+}$ or MDABCO$^{2+}$. As the frontier valence orbitals are primarily halide in character (Figure \ref{fig:DOS}), it is not surprising that the largest change in band gap is controlled by halide exchange. These findings are also supported by the DOS presented in Ref.~\citenum{Wang2019}  and are consistent with observations in hybrid halide perovskites.\cite{Butler2015}

There is a secondary contribution to the $d_{\rm norm}$ arising from the nature of the organic cation at the $A$-site. As seen in Figure~\ref{fig:NLO}, $d_{\rm norm}$ follows the order $d_{\rm norm}$(CNDABCO$^{2+}$)~$\gg$~$d_{\rm norm}$(MDABCO$^{2+}$)~$>$~$d_{\rm norm}$(ODABCO$^{2+}$). However, this trend becomes $d_{\rm norm}$(CNDABCO$^{2+}$) $\gg$ $d_{\rm norm}$(ODABCO$^{2+}$) $>$ $d_{\rm norm}$(MDABCO$^{2+}$) when the structures are fully relaxed (see Table S5). This is despite the fact that both CNDABCO$^{2+}$ and ODABCO$^{2+}$ systems have larger dielectric constants and smaller band gaps than the MDABCO$^{2+}$ counterparts. As an example, while the ODABCO$^{2+}$ cation is structurally very similar to the group MDABCO$^{2+}$ (see Figure~\ref{fig:structures}), the apical methyl group in MDABCO$^{2+}$ is substituted by a polar OH$^-$  in ODABCO$^{2+}$.  The presence of polar moieties in the organic cations, such as the CN$^-$  and OH$^-$ groups in CNDABCO$^{2+}$ and ODABCO$^{2+}$, can result in a greater intrinsic electric dipole moments, whose polarisation induces an increase of the dielectric constant.   The intrinsic electric dipole moment introduced by the CN$^-$ or the OH$^-$ groups increase the overall dipole of CNDABCO$^{2+}$ from $\sim$3.61~Debye to $\sim$7.64~Debye, compared to MDABCO$^{2+}$ $\sim$1.64 Debye, which is reflected by the values of dielectric constants of Figure~\ref{fig:NLO}. 

 Thus, the highly modular nature of the $A$ site cation, in particular the DABCO group, afforded by a metal-free scaffold can be exploited to fine-tune the NLO response of these perovskites. For example, the substitution of the CH$_3$ group on MDABCO with highly polar organic groups, such as $-$SH (thiol), $-$NH$_2$ (amine), $-$NO (nitroso) and $-$NO$_2$ (nitro), whose computed electrical dipole moment are shown in Figure~{\ref{fig:DOS}}b, represents a viable strategy to increase the NLO response of these metal-free perovskites. Indeed, in Figure~{\ref{fig:nindexes}} we have demonstrated that replacing the CH$_3$ (or OH)  with the CN moiety, increases substantially the values of d$_{\rm norm}$, thus providing a clear design rule to design novel metal-free perovskites.

The properties and tunability demonstrated in metal-free perovskites are comparable to organic-based NLO materials. Organic NLOs can supersede the performance in terms of descriptors, $\chi^{(2)}$ (see Eq.~{\ref{eq:beta}}), even by 1-2 orders of magnitude of the  inorganic NLO materials discussed so far.{\cite{Zyss1981,Eaton1991,Prasad1991,Kaino1993,Pan1996,Albert1998,Jiang1999,Kuo2001,Bosshard2002,Evans2002,Marder2006,Wang2011}} Some of these materials are even commercial.{\cite{Eaton1991,Yamada2013,NEO-823}} For example, the $4$-N,N-dimethylamino-$4'$-N$'$-methyl-stilbazolium~tosylate (DAST) forms organic  crystals  with a superior non-linear susceptibility $d$ $\sim$~580~pm/V (with $d = 0.5 \chi^{(2)}$).{\cite{Pan1996,Bosshard2002}} However, organic-based NLOs  display at least three major  limitations curbing their use in optical devices:{\cite{Jiang1999,Marder2006}} \emph{i}) difficult to obtain sufficiently large-size crystals, \emph{ii}) limited thermal stability above 200~$^{\circ}$C and \emph{iii}) low mechanical strength. For example, while a good NLO organic material, such as the 3-methyl-4-methoxy-4$'$-nitrostil-bene has a 1$^{\rm st}$ hyperpolarizability $\beta$ (see Eq.~{\ref{eq:beta}}) approximately 300 times larger than the inorganic KH$_2$PO$_4$, it melts at 109 $^{\circ}$C.{\cite{Jiang1999}}  Metal-free perovskites also offer the potential to overcome these challenges.

All of the above trends are also dependent on crystal structure and therefore a further understanding of how composition and structure are related is required to achieve truly targeted synthesis. In systems like these metal-free perovskites an interplay of weak forces, such as dispersion, hydrogen bonding and entropy will be important for driving structure and phase transitions, as exemplified in the hybrid halide perovskites.\cite{Butler2018, Kieslich2018}


\section{Conclusions}


Using first-principles calculations, we have demonstrated the existence of NLO activity in a number of novel metal-free perovskites. In the past, density functional theory has been shown to systematically underestimate the values of $\chi^{(2)}_{ijk}$. While we believe the superior quality of hybrid DFT/PBESol0-based simulations may still underestimate $\chi^{(2)}$, our data show that MDABCO-NH$_4$I$_3$ and ODABCO-NH$_4$I$_3$ both fulfil the criteria to useful as deep-UV second-harmonic generation materials. We identify the role of the dipole moment imparted by the organic group on the $A$ cation as an important parameter to tune the NLO properties of these materials. We apply this knowledge introducing the cyanide CN$^-$ group on the DABCO-NH$_4$X$_3$ structure, improving significantly the  NLO properties of metal-free perovskites. Furthermore our systematic calculations allows us to extract useful trends to chart the second-harmonic generation properties of this novel class of materials. We have shown that the material band gap, and hence the $d_{\rm norm}$ can be tuned by altering the halide anion. We also show the selection of the $A$-site cation provides an extra degree of tuning the  $\chi^{(2)}$. By combining engineering of both the $A$- and $X$-sites we believe that these findings provide a blueprint for how to achieve high non-linear optics and second-harmonic generation activity in the desired part of the electromagnetic spectrum. We hope that these findings will help to accelerate the application of metal-free perovskites as non-toxic, earth-abundant materials for the next generation of optical communication applications.


\section{Linear and non-linear optic effects}

Materials respond differently to incident electromagnetic waves, whose induced electric polarisation $P$ follows:\cite{Boyd2008} 
\begin{eqnarray}
\label{eq:polarisation}
P & = & \varepsilon_0 \left[\chi^{(1)}E + \chi^{(2)}E^2 + \dotsc + \chi^{(n)}E^n\right] \\ \nonumber
 & = & P^{(1)} + P^{(2)} + \dotsc + P^{(n)}
\end{eqnarray}
where $E$ is the electromagnetic field, $\varepsilon_0$ is the vacuum permittivity, and $\chi^{(1)}$, $\chi^{(2)}$ and $\chi^{(n)}$ are the 1$^{st}$, 2$^{nd}$ and n$^{nt}$ order electric susceptibilities. The dependence of the polarisation on the 2$^{nd}$ order  term ($P^{(2)}$) and successive electric susceptibilities suggests that the response of a permanent dipole in non-linear optical materials assumes an anharmonic behaviour once the electric dipole is perturbed by an incident electromagnetic radiation.\cite{Boyd2008}  

Here, we assess the linear and NLO properties up to $\chi^{(2)}$ in the static regime, and accurately derived from first-principles calculations using the coupled perturbed Kohn-Sham theory (CPKS).\cite{Ferrero2008,Ferrero2008a}  We compute the $n^{th}$-order derivatives of the total energy $E _{\rm Tot.}$ with respect to derivatives of the electric field components in $\vec{\epsilon _{ijk}}$ ($i$, $j$ and $k$ are the cartesian direction of the electric field), which are cast in the form of order-$n+1$ tensors. Such derivatives link to important optical descriptors, such as  the electric dipole moment $\mu _i$ and the polarisability, $\alpha_{ij}$,
\begin{equation}
\label{eq:dipole}
 \mu _i = - \frac{\partial E _{\rm Tot.} }{ \partial {\vec{\epsilon _i}}}. 
\end{equation}
\begin{equation}
\label{eq:alpha}
\alpha_{ij} = - \frac{\partial ^2 E_{\rm Tot.}}{\partial{\vec{\epsilon_i}}\partial{\vec{\epsilon_j}}}.
\end{equation}
From $\alpha_{ij} $, the components of the dielectric tensor $\varepsilon$ are derived as $\frac{\alpha_{ij}}{\varepsilon_0  V}$ 
with $\varepsilon_0$ the vacuum permittivity and $V$ the unit-cell volume. We note that $\varepsilon  = n^2  = 1 + \chi^{(1)}$, with $n$ the refractive index. Thus, $n$ and $\Delta n$ (of Eq.~\ref{eq:spreadn}) are computed directly from the dielectric tensor. 

The third-order rank tensor $\chi^{(2)}_{ijk}$ relates to 1$^{\rm st}$ hyperpolarizability $\beta_{ijk}$.
\begin{equation}
\label{eq:beta}
\beta_{ijk} = - \frac{\partial ^3 E_{\rm Tot.} }{\partial{\vec{\epsilon_i}}\partial{\vec{\epsilon_j}}\partial{\vec{\epsilon_k}} }, \; \mathrm{with} \;  \chi^{(2)}_{ijk}= \frac{2\pi \beta_{ijk}  }{V}.
\end{equation}
where $j$ and $k$ are the directions of the incident waves and $i$ the direction of the SHG wave. From Eq.~{\ref{eq:polarisation}}, $\chi^{(2)}_{ijk}$ links directly to the electric field that in materials is subjected to the spatially uniform electric fields of the incident radiations, and reads as:
\begin{equation}
\label{eq:chi2}
P_i^{(2)}(\omega_1 + \omega_2) = \varepsilon_0 \sum_{j,k} \chi^{(2)}_{ijk} \vec{\epsilon_j}(\omega_1) \vec{\epsilon_k}(\omega_2)
\end{equation}
where $\chi^{(2)}_{ijk}$ is a  component of the $\chi^{(2)}$ tensor proportional to the polarization generated along the $i$-axis (e.g., $x$) , from the $j$ and $k$ (e.g., $y$ and $z$) components of the electric fields of the incident radiations oscillating at frequencies $\omega_1$ and $\omega_2$, respectively. In SHG materials $\omega_1 = \omega_2$. 

Typically, $\chi^{(2)}_{ijk}$ is reported as $d_{ijk} = 0.5 \chi^{(2)} _{ijk}$.  Note that for hybrid functional calculations ---the choice in this study---  CRYSTAL17  computes only  $\beta_{ijk}$, $\chi^{(2)} _{ijk}$ and   $d_{ijk}$ values in the static limit.

\section{First-principles calculations}

DFT is used to investigate the structural and optical properties of the metal-free perovskites. Structure relaxations of the experimental coordinates  of the metal-free perovskites  was performed using the PBESol\cite{Perdew2008} functional as in VASP.\cite{Kresse1993,Kresse1996,Kresse1996a} The Grimme D3\cite{Grimme2010} correction was added to capture dispersive forces. The wave functions of the valence electrons were expanded with plane-waves  with a cutoff of 520 eV, and  core electrons were treated with the projector augmented wave theory.\cite{Kresse1999} The Brillouin-zone was integrated on a mesh with reciprocal density of 64 $k$-points per \AA$^{-1}$. The total energy was converged within 10$^{-5}$ eV and forces to 10$^{-2}$~eV \AA$^{-1}$. 

An accurate treatment of the electronic structure and in particular the assessment of the optical band gaps is crucial for the description of NLO properties.\cite{Lacivita2009,Rondinelli2015,Halasyamani2018} Starting from the PBEsol-D3 structures with VASP, we re-optimised the coordinates (or volume and coordinates) with the hybrid functional PBE0Sol-D3,\cite{Adamo1999,Perdew2008} including 25\% of exact exchange as available in CRYSTAL17.\cite{Dovesi2017,Dovesi2018} We exploited the all-electron linear combination of atomic orbitals of CRYSTAL17 expanded by consistent Triple-$\zeta$ plus polarisation basis-sets, see Refs. \citenum{Peintinger2012} and \citenum{Laun2018}. Given the electronic configuration of I ($\rm{[Kr]}4d^{10}5s^25p^5$) a fully-relativistic pseudo-potential is used.\cite{basisetECP} The total energy was converged within $\sim$~3$\times$10$^{-9}$ eV and integrated over a well converged and symmetrized 4$\times$4$\times$4 $k$-point mesh (i.e., the shrinking factor is set to 4). The truncation of the (infinite) Coulomb and exchange series was set by the tolerances: 10$^{-7}$, 10$^{-7}$, 10$^{-7}$, 10$^{-7}$ and 10$^{-30}$.  Table~S1 and S2 of SI show the PBE0Sol+D3 geometries.   
We computed the non-linear-optical properties with these settings.  The iterative solution of the CPKS  equations is reached for values below 10$^{-4}$. Table S6 shows a comparison of the performance of a number of DFT functionals.

The dipole moments of the DABCO units substituted by the organic groups were computed with Gaussian16\cite{g16} using the PBE0+D3\cite{Adamo1999} functional and the same basis-set used in the periodic calculations with CRYSTAL17.


\section{Competing interests}
\noindent The Authors declare no Competing Financial or Non-Financial Interests.
\section{Author contribution}
\noindent P. C.\ and K. T.\ B.\ conceived the manuscript and wrote the first draft.
\section{Acknowledgements}
\noindent P.\ C.\ acknowledges support from the Singapore Ministry of Education Academic Fund Tier 1 (R-284-000-186-133). This work benefited from access to the University of Oregon high performance computer, Talapas.
%


\bibliography{biblio}

%
\end{document}